\documentclass[11pt]{iopart}
\usepackage{iopams}
\usepackage{epsfig}
\usepackage{epstopdf}
\begin{document}
\newcommand*{\Sp}{S_L^{\mathrm {path}}}
\newcommand*{\Se}{S_L^{\mathrm {end}}}
\jl{1}
\title[Phase-space path integral for the Wigner function]{Phase-space path-integral calculation of the Wigner function}
\author{J H Samson}
\address{Department of Physics, Loughborough University, 
Loughborough, Leics LE11 3TU, United Kingdom}
\ead{j.h.samson@lboro.ac.uk}
\begin{abstract}
The Wigner function $W(q,p)$ is formulated as a phase-space path integral, whereby its sign oscillations can be seen to follow from interference between the geometrical phases of the paths.  The approach has similarities to the path-centroid method in the configuration-space path integral.  Paths can be classified by the mid-point of their ends; short paths where the mid-point is close to $(q,p)$ and which lie in regions of low energy (low $P$ function of the Hamiltonian) will dominate, and the enclosed area will determine the sign of the Wigner function.  As a demonstration, the method is applied to a sequence of density matrices interpolating between a Poissonian number distribution and a number state, each member of which can be represented exactly by a discretized path integral with a finite number of vertices.   Saddle point evaluation of these integrals recovers (up to a constant factor) the WKB approximation to the Wigner function of a number state. \end{abstract}

\pacs{03.65.Vf, 03.65.Sq, 31.15.Kb}
Published \JPA \textbf{36}, 10637--10650 (2003)
\maketitle

\section{Introduction} \label{intro}
The rapidly advancing field of quantum information requires careful characterization of quantum states.   While the state of an individual object cannot be determined, measurements on an ensemble can be used to reconstruct the state.  It is not sufficient to perform ensemble measurements on a single observable, as relative phases are unobtainable.  Rather, one requires a set of ensemble measurements on a sufficient number of non-commuting variables.  It is then possible to transform the resulting set of probabilities to obtain a complete description of the state; such a procedure is known as quantum tomography.  While the density matrix is the standard specification of a quantum state, it is often useful to represent states as quasiprobability distributions in phase space. The \emph{Wigner function} $W(q,p)$  \cite{Wig32,HOSW84} has many useful properties as a phase space distribution; in particular, its marginals give the correct probability density of commuting observables and can be used for tomographic reconstruction of the full Wigner function.  In deference to the Heisenberg uncertainty principle, the Wigner function is not a true joint probability density for position and momentum, and can be negative.  The structure shows the non-classical nature of the state; interference effects apparent in pure states are smeared out in a mixed state.  It is widely used to describe quantum states  in quantum optics \cite{Schleich,SZ} and has  been determined tomographically in a number of systems, for example helium atoms passing through double slits \cite{KPM97}, the electromagnetic field in a microwave cavity \cite{KABW98} and a Josephson junction coupled to a microwave cavity \cite{KHEVCR03}.

Expectation values take a form reminiscent of classical
statistical mechanics: they are averages with respect to a
distribution $W(q,p)$ in phase space $\Gamma$:
\begin{equation}
	\langle \hat{f} \rangle = \Tr (\hat{f}\hat{\rho}) = 
	\int_{\Gamma} 
	\rmd q\,\rmd p\; f(q,p) W(q,p).
	\label{eq:expval}
\end{equation}
Here $f$ is a function that depends only on the operator $\hat{f}$ of
interest, and the Wigner function $W$ is a normalized distribution that depends only on
the density matrix $\hat{\rho}$ \cite{Strat57}.  Although \emph{quasiprobability} distributions
$W(q,p)$ satisfying equation (\ref{eq:expval}) indeed exist \cite{HOSW84,Mueck86}, the only pure states for which the Wigner function is positive definite are Gaussian. A Gaussian smoothing over a phase space area of $\frac 1 2 \hbar$ results in the positive-definite Husimi (or $Q$) function.   However, the integral of the Wigner function over a bounded region in phase space may still be negative even if its area is larger than the uncertainty bound $\frac 1 2 \hbar$ \cite{BDW99}.  Hence the approach to the classical limit, as a delta function on the classical torus, is highly singular \cite{Berry77}.

Path integration \cite{FH} allows evaluation of expectation values in a formally similar way to equation (\ref{eq:expval}):
\begin{equation}
	\langle \hat{f} \rangle = \Tr (\hat{f}\hat{\rho}) = 
	\int_{\mathcal P} 
	\mathcal D q\,D p\;f[q,p] e^{-S[q,p]}.
	\label{eq:expvalpi}
\end{equation}
The integration is now over \emph{paths} $(q(\tau),p(\tau))$ in phase space.  The distribution of paths $e^{-S[q,p]}$ is now in general complex, as the action $S[q,p]$ carries a geometric phase factor.

The purpose of this work is to relate these two formulations of a distribution, and show how the phases in the path integral contribute to the non-positivity of the Wigner function.  To do this we could associate a subset of the paths to each point in phase space.  This recalls the path-centroid method: integration of all non-zero frequency modes out of a configuration-space path integral maps the system to an effective \emph{classical} system, 
usually determined variationally
\cite{FH,GT85,FK86}.  Paths ${\bi x}(\tau), 0 \le \tau < \beta$ are
classified according to their \emph{path centroid} ${\bar{\bi x}} =
\frac{1}{\beta} \int_{0}^{\beta}{\bi x}(\tau){\mathrm d}\tau$.  The
path integral with action ${\cal S}[\bi{x}]$ reduces to an ordinary
integral over $c$-number variables with classical effective
Hamiltonian $H_{\mathrm{eff}}(\bar{\bi x})$.  Excursions of the path
from $\bar{\bi x}$ provide quantum corrections to the potential in the
effective Hamiltonian.  The function $W(\bar{\bi x})=\exp(-\beta
H_{\mathrm{eff}}(\bar{\bi x}))$, regarded as a Boltzmann
distribution of the variables $\bar{\bi x}$, forms the basis of
a classical statistical mechanics in phase space \cite{CV94}.  This distribution is an intuitive
interpretation of the path centroid distribution \emph{if the
observables represented by $\bi{x}$ are compatible}, as is indeed the case for configuration-space path integrals, and has successfully dealt with quantum corrections to the dynamics of solids.  However, this is
not the case when the variables are incompatible: the action in the path integral is complex, and the resulting phase space distribution is not positive definite.

The present author has previously shown a correspondence 
between the path-centroid distribution
and the Wigner function $W(\bi S)$ for components of a free spin $s$ \cite{S23, S32}.  This is a singular distribution, with derivatives of delta functions supported on spheres of quantized radius.    The distribution of the path centroids of spherical polygons of $L$ vertices  on the Bloch sphere converges as $L\rightarrow \infty$ to the Wigner function, where the weight of a polygon of area $\Omega$ includes a Berry phase $s\Omega$.  (We consider the Bloch sphere as embedded in $\mathbb R^3$.)  For a given value of the path centroid inside the Bloch sphere, the dominant path is a small circle enclosing non-zero area and therefore contributing a phase to the Wigner function.  More generally, the Wigner function of a set of operators can be computed
as a histogram of their path averages:
\begin{equation}
	W({\bi x})\equiv\langle\delta_{N}({\bi x}-\hat{\bi x})\rangle=
	\lim_{L\rightarrow\infty}\langle\delta_{N}({\bi
	x}-\bar{\bi P})\rangle_{L}.
	\label{eq:result}
\end{equation}
Here $\hat{\bi x}=(\hat x_1,\ldots,\hat x_N)$ are operators commuting with the Hamiltonian, $\bi x \in \mathbb R^N$ are $c$-number variables and the $P$ functions of the operators are defined as distributions over the coherent state manifold $\Gamma$ (e.~g., the Bloch sphere $S^2$):
\begin{equation}
\hat x_i=\int_\Gamma \rmd \mu(\gamma) P_i(\gamma) |\gamma\rangle \langle\gamma|, i=1\ldots N.  
\end{equation}

\begin{equation}
\bar P_i = \int_0^1 P_i(\gamma(\tau))\rmd \tau
\label{eq:Pbar}
\end{equation}
is the path centroid, the time average of the $P$ function.  The first set of angle brackets in equation (\ref{eq:result}) represents a
thermal expectation value, and the second an average over $L$-vertex 
polygonal paths in the
coherent-state path integral; $\delta_{N}$ is the symmetrically-ordered delta function in 
${\mathbb R}^{N}$ (see equation (\ref{eq:deltaW}) below). 

The above result is only applicable to compact phase space, where the excursions of the path are bounded for a free particle.  The sign oscillations of the Wigner function are a consequence of the geometrical phase arising from the curvature of the phase space (or coherent-state manifold). In the present work we obtain an equivalent result for flat phase space, with a Hamiltonian confining the dynamics to a region of phase space.  The Wigner function is no longer the distribution of the path centroid, but is linked to a distribution of the midpoint of the ends of the path.  The paths are confined to regions where the Hamiltonian is small, and the sign oscillations of the Wigner function are related to the symplectic area enclosed by the path.

In the next section we review the relevant properties of coherent states and phase space distributions.  Section \ref{sec:calc} contains the main result, a path integral expression for the Wigner function (\ref{eq:WignerL2}--\ref{eq:Sprime}).  We apply this to obtain the Wigner function for a number state in Section \ref{sec:number}.  By saddle-point evaluation of the path integral we recover the usual WKB approximation to the Wigner function.  Section \ref{sec:conc} contains some concluding remarks.

\section{Phase space distribution}
\label{sec:def}
\subsection{Coherent states}

We consider a two-dimensional phase space $\Gamma = \mathbb R^2$ with a single coordinate $q$ and conjugate momentum $p$. 
The Hilbert space $\mathcal{H}$ is spanned by the number states $\{|n\rangle, n = 0, 1, \ldots\}$ of a harmonic oscillator of (arbitrary) mass $m$ and natural frequency $\omega$.
The raising and lowering operators
\begin{equation}
\label{eq:ladder}
\hat a^\dag = \sqrt{\frac{m\omega}{2\hbar}}\hat q - \frac{\rmi}{\sqrt{2m\hbar\omega}}\hat p
\:\mathrm{~and~} \:
\hat a = \sqrt{\frac{m\omega}{2\hbar}}\hat q + \frac{\rmi}{\sqrt{2m\hbar\omega}}\hat p 
\end{equation}
satisfy the canonical commutation relations
\begin{equation}
\label{eq:comm}
[\hat a, \hat a^\dag] = 1.
\end{equation}

\emph{Coherent states} $|\alpha\rangle\in\mathcal H$, with complex label $\alpha \in \mathbb C$ \cite{SZ}, provide a continuous basis for Hilbert space: 
\begin{equation}
\label{eq:cs}
|\alpha\rangle \equiv e^{\alpha \hat a^\dag-\alpha^* \hat a}|0\rangle = 
\rme^{-|\alpha|^2/2} \sum_{n=0}^\infty \frac {\alpha^n}{\sqrt{n!}}|n\rangle.
\end{equation}
(In this paper, kets with Roman or numerical indices will always refer to harmonic oscillator basis states, and kets with Greek indices will always refer to coherent states.)  Coherent states are normalized eigenkets of the lowering operator, $\hat a|\alpha\rangle = \alpha|\alpha\rangle$, but are not orthogonal:
\begin{equation}
\label{eq:overlap}
\langle \beta | \gamma \rangle = e^{-\frac 1 2 |\beta|^2-\frac 1 2 |\gamma|^2 +\beta^*\gamma}.
\end{equation}
They form an overcomplete basis for Hilbert space:
\begin{equation}
\label{eq:complete}
\frac 1 \pi  \int \rmd^2\alpha |\alpha\rangle \langle\alpha| =1,
\end{equation}
where $\int \rmd^2\alpha = \int_{-\infty}^{\infty} \rmd\,\mathrm{Re}\alpha \int_{-\infty}^{\infty} \rmd\,\mathrm{Im}\alpha$.
We note the one-to-one correspondence between the complex plane and phase space,
\begin{equation}
\alpha = \sqrt{\frac{m\omega}{2\hbar}} q + \frac{\rmi}{\sqrt{2m\hbar\omega}} p;
\label{eq:complex}
\end{equation}
the ket $|\alpha\rangle$ is a minimum-uncertainty state centred at the point $(q,p)$.

\subsection{The $P$ function}
The quantum state of a system is specified by its density matrix
$\hat{\rho}$, a positive Hermitian operator
with unit trace.   This can be written as a weighted average of coherent-state projections,
\begin{equation}
\label{eq:P}
\hat \rho = \int \rmd^2\alpha\, P(\alpha,\alpha^*) |\alpha\rangle \langle\alpha|.
\end{equation}
The  (Glauber-Sudarshan) $P$ function, as implicitly defined above, (see e.~g.~references \cite{SZ,CG69a,CG69b}) is the expectation value of the normal-ordered $\delta$ operator:
\begin{equation}
\label{ eq:Pdelta}
P(\alpha,\alpha^*) = \Tr \left[\hat\rho \delta(\alpha^*-\hat a^\dag)\delta(\alpha-\hat a)\right],
\end{equation}
where
 \begin{equation}
		\delta(\alpha^*-\hat a^\dag)\delta(\alpha-\hat a) \equiv 
		\int\frac{{\rmd}^{2}c}{\pi^{2}}
		\rme^{\rmi c(\alpha^*-\hat a^\dag)}e^{ \rmi c^*(\alpha-\hat a)}.
    	\label{eq:deltaP}
    \end{equation}

Clearly (\ref{eq:P}) the $P$ function of a coherent state is a delta function.  For number states it is a more singular distribution.

\subsection{The Wigner function}
Any classical state is fully specified by a non-negative distribution $W_{\mathrm{cl}}(q,p)$ over phase space  such that the expectation value of a function $f(q,p)$ is
\begin{equation}
\langle f(q,p) \rangle_{\mathrm{cl}} =  
	\int_{\Gamma} 
	\rmd q\,\rmd p\; f(q,p) W_{\mathrm{cl}} (q,p).
\label{eq:cl}
\end{equation}
The expectation value of an operator $\hat A$ can nevertheless still be written in the above form (\ref{eq:cl}),
 \begin{equation}
\label{eq:qm}
\Tr (\hat f \hat \rho) =  
	\int_{\Gamma} 
	\rmd q\,\rmd p\; f(q,p) W (q,p),
\end{equation}
where the $c$-number symbols $f(q,p)$ and $W (q,p)$ are linear functions of the operators $\hat f$ and $\hat \rho$ respectively.  There is a one-dimensional family of such functions, which includes the $P$, Wigner and $Q$ functions, corresponding to distributions for normal, symmetrical and anti-normal ordering of the operators \cite{CG69a,CG69b,BM99}. 
We shall compute the Wigner function $W(q,p)$ \cite{Wig32,HOSW84}, which we shall subsequently write as a function in the complex plane.    $W(\alpha,\alpha^*)$ is a linear function of $\hat{\rho}$, 
 \begin{equation}
  	W(\alpha,\alpha^*) = \Tr \left[\hat{\rho}\, 
  	  	  	\delta_{2}(\alpha-\hat{a})\right] ,
  	\label{eq:Weyl}
  \end{equation}
  where $\delta_{2}$ is the symmetrically ordered $2$-dimensional delta function: 
  \begin{equation}
		\delta_{2}(\alpha-\hat{a}) \equiv 
		\int\frac{\rmd^2c}{\pi^{2}}
		\rme^{\rmi c(\alpha^*-\hat a^\dag) + \rmi c^*(\alpha-\hat a)}.
    	\label{eq:deltaW}
    \end{equation}
To evaluate the matrix elements of this operator, we use the overlap (\ref{eq:overlap}) and the Baker-Campbell-Hausdorff identity $\rme^{\hat A+\hat B} = \rme^{-[\hat A,\hat B]/2}\rme^{\hat A} \rme^{\hat B}$ (for the case where $\hat A$ and $\hat B$ commute with their commutator):
\begin{equation}
\label{eq:BCH }
\rme^{-\rmi c\hat a^\dag - \rmi c^*\hat a} = \rme^{-|c|^2/2}\rme^{-\rmi c\hat a^\dag}\rme^{- \rmi c^*\hat a},
\end{equation}
giving
\begin{eqnarray}
\left\langle \beta \left|\frac \pi 2 \delta_2(\alpha-\hat{a}) \right| \gamma\right\rangle & = &
\int\frac{\rmd^2 c}{2\pi} \rme^{\rmi c\alpha^*+\rmi c^*\alpha-|c|^2/2}\langle\beta |\rme^{-\rmi c\hat{a}^\dag} \rme^{-\rmi c^*\hat{a}}|\gamma\rangle
\\ & = &  \int\frac{\rmd^2c}{2\pi} \rme^{\rmi c\alpha^*+\rmi c^*\alpha-|c|^2/2}\rme^{-\rmi c\beta^*-\rmi c^*\gamma}\langle \beta |\gamma \rangle \\
& = & \rme^{-2(\alpha-\gamma)(\alpha^*-\beta^*)}\langle \beta |\gamma\rangle.  \label{eq:disppar3}
\end{eqnarray}
The delta operator is also a  ``displaced parity" operator \cite{BV94}, which inverts a coherent state about the point $\alpha$:
\begin{equation}
\label{eq:disppar}
\left\langle \beta \left|\frac \pi 2 \delta_2(\alpha-\hat{a}) \right| \gamma\right\rangle = 
\rme^{-\alpha\gamma^*+\alpha^*\gamma} \langle \beta | 2\alpha-\gamma \rangle.
\end{equation} 
The pre-factor here is a geometrical phase factor, with phase four times the area of the triangle $0\alpha\gamma$ (or the area of the triangle with mid-points $0,\alpha,\gamma$ \cite{OdA92,OdA98}).

The Wigner function is a Gaussian convolution of the $P$ function of the density matrix,
\begin{equation}
\label{eq:WP}
W(\alpha,\alpha^*) = \frac 2 \pi \int \rmd^2\beta \rme^{-2|\alpha-\beta|^2} P(\beta,\beta^*).\end{equation}
The Wigner function exists and is bounded for any density matrix, even if the $P$ function is unbounded.  Its marginals, obtained by integrating over one variable, are the probability densities of the conjugate variable; the Wigner function can in turn be reconstructed from the ensemble of measured marginals.
 
\section{Path integral calculation of Wigner function}\label{sec:calc}
Path integration \cite{FH,Kleinert} is an important technique both analytically and, increasingly, numerically for the evaluation of a thermal expectation value as an integral over paths.  If these paths are sequences of points in a state space they will acquire a geometrical phase (even in an imaginary-time path integral).  Thus the evaluation of the integral over paths will rely on the interference between complex weights; in the numerical context the difficulty in evaluating such a integral by random sampling constitutes the Monte Carlo sign problem \cite{S23}.  Formally we can write the path integral in a similar form to equation  (\ref{eq:qm}), where the integration is over paths $\gamma(\tau)$ in the path space $\mathcal P: [0,1)\rightarrow \Gamma$:  
\begin{equation}
\label{eq:qmpath}
\Tr (\hat f \hat \rho) =  
	\int_{\mathcal P} 
	\mathcal{D}\gamma f[\gamma] \rho[\gamma],
\end{equation}
where $f[\gamma]$ and $\rho[\gamma]$ are functionals defined in terms of $\hat f$ and $\hat \rho$.  

To relate this to our discussion of Wigner functions as distributions over phase space, we recall the path-centroid method for obtaining effective potentials in configuration-space path integrals \cite{FH,GT85,FK86,CV94}.  Here each path $q(\tau)$ is associated with the path centroid  $\bar q = \int_0^1 q(\tau)d\tau$ and the path integral is performed over all paths $q(\tau)$ subject to this constraint.  The resulting distribution of  $\bar q$, interpreted as a Boltzmann distribution $\exp(-\beta V_\mathrm{eff}(\bar q))$, allows definition of an effective potential $V_\mathrm{eff}$.  

More generally, we can classify paths by points in phase space, $C: \mathcal P \rightarrow \Gamma$, where $C$ is not necessarily the path centroid.   The appropriate classification, we will see later, is the centre of the chord linking the end-points of the path, $C[\gamma]=(\gamma(0)+\gamma(1))/2$.  Let $\mathcal P_{\alpha}\equiv\{\gamma \in \mathcal P:  C[\gamma] = \alpha\}$ be the space of paths associated with a point $ \alpha \in \Gamma$ and assume that the path integral in $\mathcal P_{\alpha}$ can be performed in some approximation, leaving a two-dimensional integral over phase space and mapping functionals of paths onto functions in phase space:  
\begin{equation}
\label{eq:qmclass1}
\int_{\mathcal P} 
	\mathcal{D}\gamma f[\gamma] =  
	\int_{\Gamma} 
	\rmd^2\alpha \tilde f(\alpha) 
\end{equation}
where
\begin{equation}
\label{eq:qmclass2}
\tilde f(\alpha) = \int_{\mathcal P_{\alpha}} 
	\mathcal{D}\gamma f[\gamma] = \int_{\mathcal P} 
	\mathcal{D}\gamma f[\gamma] \delta_2(\alpha - C[\gamma]).
\end{equation}
The contribution of the subspaces in this integral can then be interpreted as a quasidistribution over phase space; any non-positivity arises from interference between the geometrical phases of the paths, the sign relating to the distribution of areas of paths in $\mathcal P_{\gamma}$.  Of course, the integrand $A[\gamma]\rho[\gamma]$ in the path integral (\ref{eq:qmpath}) will not in general separate after path classification.

The relation of the phase of the Wigner function to areas in phase space is  not new, as Schleich has extensively reviewed \cite{Schleich}.  Berry \cite{Berry77} has established a relationship between the Wigner function and areas in phase space in a semiclassical context: a line segment is drawn that intersects the classical torus in two points equidistant from $(q,p)$, and the phase is then the area enclosed by the line segment and the classical torus.  Ozorio de Almeida \cite{OdA92} has obtained a path integral representation for the Wigner function of a product of operators, and hence for the evolution operator, in terms of the star product of Wigner functions.  Here the geometric factor is the symplectic area of a polygon whose mid-points are the phase space points on the path.

Our approach is complementary to and simpler than that in references \cite{OdA92}, and relates the Wigner function to a path integral of a product of $P$ functions.  There is a choice of ways of evaluating a coherent-state path integral, using different symbols for the operators concerned \cite{VS94}.  The present approach starts from the usual time-slicing of the density matrix.  Suppose we wish to find the density matrix $\hat\rho=\exp(-\beta\hat H)$ of a Hamiltonian whose $P$ function $P_H(\alpha,\alpha^*)$ exists.  This class includes (but is not restricted to) Hamiltonians polynomial in $\hat a$ and $\hat a^\dag$ \cite{CG69a,CG69b}.  The density matrix can be time sliced as
\begin{equation}
\label{eq:Trotter}
\hat \rho =  \lim_{L\rightarrow\infty} \left(1-\beta \hat H/L \right)^L.
\end{equation}
(There is no additional difficulty in introducing a Trotter decomposition of the Hamiltonian into non-commuting terms.)  Replacing the Hamiltonian by its $P$ representation (\ref{eq:P}) and re-exponentiation defines the $L$th approximant to the density matrix as
\begin{eqnarray}
\label{eq:TrotterL}
\hat \rho_L(\beta) &=&  \prod_{l=1}^L \int \rmd^2\gamma_l \rme^{-\beta P_H(\gamma_l,\gamma_l^*)/L} |\gamma_l\rangle \langle\gamma_l| \\
&=& (\hat \rho_1(\beta/L))^L.
\end{eqnarray}
It is useful to ensure that the approximant obtained numerically from finite time-slicing avoids unphysical pathologies \cite{S33}.  Since each approximant is (up to normalization) a physically realizable density matrix, the Wigner function is the limit of a sequence of physical Wigner functions.

\begin{figure}[t]\begin{center}
\includegraphics[width=8cm]{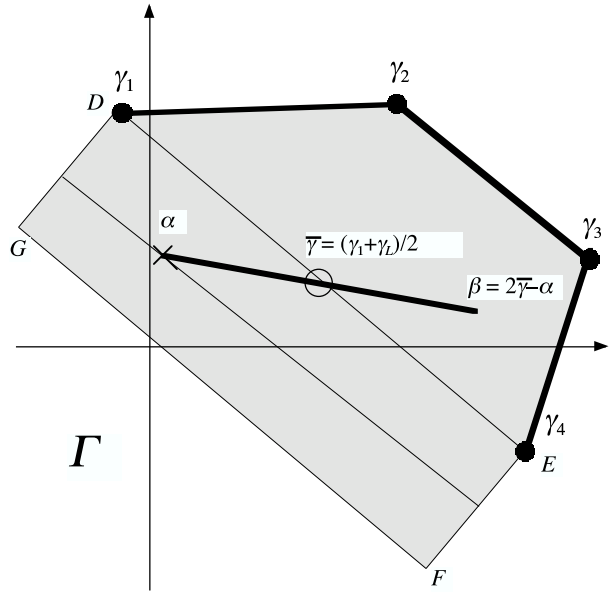} 
\end{center}
\caption{Geometrical properties of path for evaluation of Wigner function at $\alpha$ (for $L=4$).  The real part of the geometrical action depends on the sum of squares of the bold line segments; $\bar\gamma$ is the mid-point of the line  $\gamma_1\gamma_L$ joining the ends of the path, and is the mid-point of the line $\alpha \beta$.  The imaginary part is proportional to the shaded area, where $\alpha$ lies on the central line of the rectangle $DEFG$.
\label{fig:path}}
\end{figure}

The Wigner function, following Eqs.~(\ref{eq:Weyl}) and (\ref{eq:disppar3}), is
\begin{eqnarray}
\label{eq:WignerL1}
\fl W_L(\alpha,\alpha^*) = \Tr \int \prod_{l=1}^L \rmd^2\gamma_l \delta_{2}(\alpha-\hat{a})|\gamma_L\rangle\langle\gamma_{L}|\gamma_{L-1}\rangle\cdots\langle\gamma_2|\gamma_1\rangle\langle\gamma_1 | e^{-\beta \sum_{l=1}^L P_H(\gamma_l,\gamma_l^*)/L} \\
\label{eq:WignerL2}
\lo= \int \prod_{l=1}^L \rmd^2\gamma_l e^{-S_L[\gamma,\alpha] -\beta \sum_{l=1}^L P_H(\gamma_l,\gamma_l^*)/L},
\end{eqnarray}
where the geometrical action $S_L[\gamma,\alpha]=\Sp[\gamma]+\Se[\gamma,\alpha]$ has two terms: the Bargmann invariant $\Sp$ depends only on the geometrical properties of the path,
\begin{equation}
e^{-\Sp[\gamma]} =\langle\gamma_{L}|\gamma_{L-1}\rangle\cdots\langle\gamma_2|\gamma_1\rangle \langle\gamma_1|\gamma_L\rangle,
\end{equation}
\begin{equation}
\label{eq:SL}
\Sp[\gamma] = \sum_{l=1}^L \left(\frac 1 2 |\gamma_l-\gamma_{l-1}|^2 + \rmi \mathrm{Im} \gamma_l \gamma^*_{l-1} \right),
\end{equation}
and $\Se$, connecting the end-points of the path to $\alpha$, is obtained from \eref{eq:disppar3} as
\begin{equation}
\label{eq:Sprime}
\Se[\gamma,\alpha] = 2(\alpha-\gamma_L)(\alpha^*-\gamma_1^*).
\end{equation}
(\Eref{eq:SL}  uses  ``periodic boundary conditions"  $\gamma_0\equiv\gamma_L$ for notational convenience.) The total geometrical action can then be manipulated into the form
\begin{equation}
\label{eq:ReS}
\mathrm{Re}\;S_L[\gamma,\alpha]= \frac 1 2 \sum_{l=2}^L \left|\gamma_l-\gamma_{l-1}\right|^2 +
2\left|\alpha-\frac {\gamma_1+\gamma_L} 2\right|^2,
\end{equation}
\begin{equation}
\label{eq:ImS}
\mathrm{Im}\;S_L[\gamma,\alpha]= \mathrm{Im}\sum_{l=1}^L \gamma_l \gamma^*_{l-1} +
2\;\mathrm{Im}(\alpha-\gamma_L)(\alpha^*-\gamma_1^*).
\end{equation}
Figure \ref{fig:path} shows the geometrical interpretation of the above action.  The path is classified by the mean of its end-points,
\begin{equation}
\label{eq:mid}
\bar\gamma=C[\gamma] \equiv \frac {\gamma_1+\gamma_L} 2.
\end{equation}
The real part of the geometrical action is equal to half the sum of the squares of the line segments shown in bold in figure \ref{fig:path}: that is, the $L-1$ ``internal" segments of the path (excluding $\gamma_L$ to $\gamma_1$) and the line from $\alpha$ to $2\bar\gamma-\alpha$, the first and second terms in \eref{eq:ReS} respectively.  The second term suggests a Gaussian smoothing of the distribution of chord midpoints: the centre of the path end-points is close to $\alpha$, and the delta function in \eref{eq:qmclass2} is effectively replaced by a Gaussian. The dynamical part of the action favours regions of phase space where the $P$ function of the Hamiltonian is small. The imaginary part is twice the area shown shaded in the figure: the area enclosed by the path plus the area of the rectangle with the line segment $\gamma_1$ to $\gamma_L$ forming one side and $\alpha$ on the line of symmetry parallel to this side.  Since the time-reversed path changes the sign of the area but leaves the lengths invariant the Wigner function is real. Furthermore, for $\alpha$ outside the energy surface, the dominant path has all points clustered beneath $\alpha$ and encloses zero area.  The Wigner function is therefore oscillatory inside the energy surface and monotonically decaying outside. The real part of the action also represents a ferromagnetic XY model on a one-dimensional chain with $L$ spins, a local anisotropic potential $P_H/L$ allowing for amplitude fluctuations, an antiferromagnetic bond between the ends $\gamma_1,\gamma_L$ of the chain, and an external field (due to $\alpha$) acting on the end spins.

\section{Example: number state}\label{sec:number}
As a concrete example, we evaluate the Wigner function of a number state.  To do this, we define a family of density matrices $\hat\rho_L(N)$ interpolating between a Poisson state $\hat\rho_1(N)$ of mean occupation $N$ and a number state $\hat\rho_\infty(N)$.  Here $L$ is a positive integer and $N$ is a non-negative real number.  We first define $\hat\rho_1(N)$ to be a phase-randomized coherent state of magnitude $|\alpha|=\sqrt N$ \cite{Vou94}:
\begin{equation}
\label{eq:rho1}
\hat\rho_1(N) = \int_0^{2\pi} \frac {\rmd\theta}{2\pi} 
\left|\sqrt N \rme^{\rmi\theta}\right\rangle \left\langle \sqrt N \rme^{\rmi\theta}\right|
\end{equation}
From equation (\ref{eq:cs}) this is easily shown to be a Poissonian mixture of number states of mean occupation $N$:
\begin{eqnarray}
\hat\rho_1(N) & = & \int_0^{2\pi} \frac {\rmd\theta}{2\pi} 
\rme^{-N} \sum_{n=0}^\infty\sum_{m=0}^\infty \frac {N^{n/2}\rme^{in\theta}}{\sqrt{n!}}\frac{N^{m/2}\rme^{-\rmi m\theta}}{\sqrt{m!}}|n\rangle\langle m| \\
 & = & \rme^{-N} \sum_{n=0}^\infty \frac {N^n}{n!}|n\rangle\langle n|. \label{eq:Poisson}
\end{eqnarray}
The $P$ function of the distribution (\ref{eq:Poisson}) is localized on a circle,
\begin{equation}
\label{eq:P1}
P_1(\alpha,\alpha^*;N) = 2\delta(|\alpha|^2-N),
\end{equation}
and the Wigner function, obtained from the integral (\ref{eq:WP}), is
\begin{equation}
\label{eq:W1}
W_1(\alpha,\alpha^*;N) = \frac 2 \pi \rme^{-2|\alpha|^2-2N} I_0(4\sqrt N |\alpha|),
\end{equation}
where $I_\nu$ is the modified Bessel function.  The $P$ function and Wigner function are non-negative for this state.

The family of density matrices is then defined by radially squeezing the above state, reducing the number fluctuations:
\begin{eqnarray}
\label{eq:rhoL}
\hat \rho_L(N) &=& \frac {(\hat \rho_1(N))^L}{Z_L(N)}  = \frac {\rme^{-LN} }{Z_L(N)}
\sum_{n=0}^\infty \left(\frac {N^n}{n!}\right)^L|n\rangle\langle n| \\
&=& \frac 1 {Z_L(N)}\prod_{l=1}^L\left(\int _0^{2\pi} \frac {\rmd\theta_l}{2\pi} \right)
|\gamma_L\rangle\langle\gamma_{L}|\gamma_{L-1}\rangle\cdots\langle\gamma_2|\gamma_1\rangle\langle\gamma_1 | \label{eq:rhoL2}
\end{eqnarray}
with normalization factor
\begin{equation}
\label{eq:ZL}
 Z_L(N)= \Tr (\hat \rho_1(N))^L =
e^{-LN} \sum_{n=0}^\infty \left(\frac {N^n}{n!}\right)^L.
\end{equation}
The coefficient of $|n\rangle\langle n|$ dominates in the sum (\ref{eq:rhoL}) if $n<N<n+1$.   Thus, for any non-integer $N$, $\hat\rho_L(N)$ converges onto the number state $|n\rangle\langle n|$ in the limit $L\rightarrow\infty$, where $n$ is the largest integer less than $N$.  We choose $N=n+\frac 1 2$ to obtain the fastest convergence onto the state $|n\rangle\langle n|$.  This corresponds to the classical energy surface at the energy eigenvalue,
\begin{equation}
\label{eq:orbit}
 \frac {p^2}{2m} + \frac {m\omega^2x^2}{2} = \hbar\omega \left(n+\frac 1 2\right).
\end{equation}
The $P$ function of the number state is singular, while the Wigner function can be expressed in terms of Laguerre polynomials \cite{SZ,CG69b}:
\begin{eqnarray}
P_\infty(\alpha,\alpha^*;N) & = & \frac {e^{|\alpha|^2} }{n!} \frac {\partial^{2n}} {\partial\alpha^n\partial\alpha^{*n}}\delta_2(\alpha),\label{eq:Pinfty}\\
W_\infty(\alpha,\alpha^*;N) & = & \frac 2 \pi (-1)^n \rme^{-2|\alpha|^2} L_n(4|\alpha|^2).\label{eq:Winfty}
\end{eqnarray} 
Thus the $P$ symbol becomes increasingly singular with $L$, while the Wigner function remains bounded but develops sign oscillations.

It is useful to ensure that the approximants (\ref{eq:WignerL2}) to the number state Wigner function refer to physically realizable states \cite{S33}.  This is indeed the case here, as the density matrices (\ref{eq:rhoL}) constitute a sequence of thermal density matrices for the (non-polynomial) Hamiltonian
\begin{eqnarray}
\label{eq:HL}
\hat H(N) &=& N+\ln \hat n!-\hat n\ln N \\ 
\label{eq:HL2}
&=&  N+\ln \int_0^\infty e^{-x+\hat n\ln (x/N)} dx
\end{eqnarray}
at temperatures $kT = 1/L, L=1,\ldots,\infty$. The path integrals to be obtained are therefore exact expressions for the thermal Wigner function of a physically admissible Hamiltonian.

While the Hamiltonian (\ref{eq:HL2}) does not have an obvious physical interpretation, for large $N$ we can use Stirling's approximation to expand about the minimum to give a Hamiltonian, quadratic in the number operator, that might for example describe the charging energy of a Cooper-pair box
\begin{equation}
\label{eq:Hblock}
\hat H(N) \approx \frac 1 2 \ln 2\pi N + \frac {(\hat n + \frac 1 2 - N)^2}{2N}.
\end{equation}

We now calculate the path integral as defined in section \ref{sec:calc}. 
For convenience we will use polar coordinates
\begin{equation}
\label{eq:alpha}
\alpha = s \rme^{\rmi\phi}.
\end{equation}
The path vertices are restricted to the circle
\begin{equation}
\label{eq:gammal}
\gamma_l = r \rme^{\rmi\theta_l} 
\end{equation}
where $r=\sqrt N$.  The symmetry allows us to take $\phi=0$ in the following (or to define the $\theta_l$ relative to $\phi$).
The Wigner functions  (\ref{eq:WignerL2}) simplify to
\begin{equation}
\label{eq:WLexact}
W_L(\alpha,\alpha^*;N) =\frac 1 {Z_L(N)}\prod_{l=1}^L\left(\int _0^{2\pi} \frac {\rmd\theta_l}{2\pi} \right) \rme^{-S_L[\gamma,\alpha]}
\end{equation}
with the action defined as in equations (\ref{eq:SL}--\ref{eq:ImS}):
\begin{equation}
\label{eq:SN}
\fl S_L[\gamma,\alpha]= Lr^2+2s^2-r^2\sum_{l=1}^L \rme^{\rmi(\theta_{l-1}-\theta_{l})} + 
2r^2 \rme^{\rmi(\theta_{L}-\theta_{1})}-2rs(\rme^{-\rmi\theta_1}+\rme^{\rmi\theta_L}).
\end{equation}
The real part of the action is that of a ferromagnetic XY model on an open chain of $L$ spins, with an antiferromagnetic bond between the end spins and an external field acting on the end spins.  The imaginary part is related to the area enclosed.  

The saddle-point approximation to the integral (\ref{eq:WLexact}) is given in the appendix.  The result is a Wigner function  oscillatory inside the energy surface $|\alpha|=\sqrt{n+1/2}$ and decaying outside:
\begin{equation}
\label{eq:Wsp}
\fl W_\infty^{\mathrm {sp}}(\alpha,\alpha^*;n+\frac 12) \propto 
\left\{ \begin{array}{ll}
  \frac 
{\cos\left[(2n+1)\cos^{-1} \left(\frac{|\alpha|}{\sqrt{n+\frac 12}}\right)-2|\alpha|\sqrt{n+\frac 12 - |\alpha|^2}- \pi/4\right]}
{\left[\frac{|\alpha|^2}{n+\frac 1 2}\left(1-\frac{|\alpha|^2}{n+\frac 1 2}\right)\right]^{1/4}} ,   &  |\alpha|<\sqrt{n+\frac 1 2}  \\ & \\
\frac{\exp\left[(2n+1)\cosh^{-1} \left(\frac{|\alpha|}{\sqrt{n+\frac12}}\right)-2|\alpha|\sqrt{ |\alpha|^2-n-\frac 12}\right]} 
{2{\left[\frac{|\alpha|^2}{n+\frac 1 2}\left(\frac{|\alpha|^2}{n+\frac 1 2}-1\right)\right]^{1/4}}} ,   &  |\alpha|>\sqrt{n+\frac 1 2} 
\end{array}\right.
.
\end{equation}
The functional form is identical to the Wigner function evaluated from the WKB approximation to the wave function of $|n\rangle$  \cite{Berry77},
\begin{equation}
\label{eq:WWKB}
\fl W^{\mathrm{WKB}}(\alpha,\alpha^*) \approx \frac 
{\cos\left((2n+1)\cos^{-1} \left(\frac{|\alpha|}{\sqrt{n+\frac 12}}\right)-2|\alpha|\sqrt{n+\frac 12 - |\alpha|^2}- \pi/4\right)}
{(\pi^3/2)^{1/2}\left(|\alpha|^2(n+1/2-|\alpha|^2\right)^{1/4}},
\end{equation}
although the saddle-point approximation does not yield the correct numerical coefficient.

\begin{figure}[t]\begin{center}
\includegraphics[width=15cm]{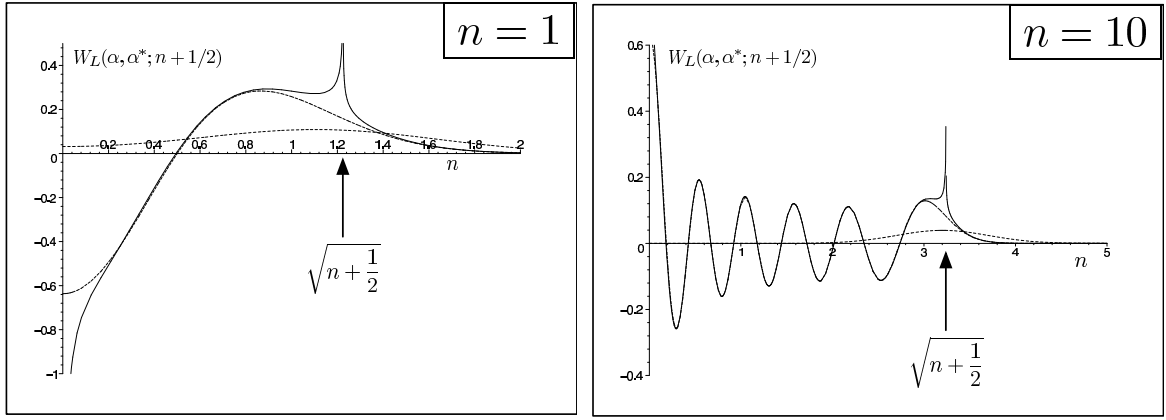} 
\end{center}
\caption{The Wigner function for the number states $|1\rangle$ and $|10\rangle$.  The dashed line represents the exact Wigner function (\ref{eq:WLexact}).  The bold line represents the saddle-point approximation derived here (\ref{eq:Wsp}); the overall amplitude has been adjusted to agree with Berry's result \cite{Berry77}.  The non-negative dotted line represents the Wigner function for the Poisson state $\hat\rho_1(n+\frac 12)$ (\ref{eq:W1}).
\label{fig:functions}}
\end{figure}

Figure \ref{fig:functions} shows the Wigner function and its saddle-point approximation; the only adjustable parameter is the overall amplitude, which is taken from equation (\ref{eq:WWKB}).  The fit is very good apart from singularities at the origin and on the energy surface.  The fit to the Wigner function for $|n\rangle$ requires fixing the parameter $N=n+\frac 1 2$.  An exact evaluation of the path integral for $W_L(\alpha,\alpha^*;N)$ should converge to the same Wigner function (\ref{eq:Winfty}) for $n<N<n+1$. The exact Wigner function  $W_\infty(\alpha,\alpha^*;N)$, must have discontinuities at every integer value of $N$.  The saddle-point approximation smears out these discontinuities, so that the Wigner function continuously interpolates between that for the number states $|n-1\rangle$ and $|n\rangle$.

\section{Conclusions}
\label{sec:conc}
The path integral formalism developed here has two purposes.  Firstly, it provides a possible method for the calculation of the Wigner function for a wide range of density matrices in phase space.  This includes, but is not restricted to, thermal distributions resulting from a polynomial Hamiltonian; a simple example might be a Cooper-pair box.  The example of a number state shows that good results can be obtained in a saddle-point approximation.  

Secondly, it provides a further geometric interpretation of the sign oscillations of the Wigner function in terms of the area enclosed by the dominant paths associated with a point $(q,p)$ in phase space.  To find the quasiprobability density that a particle is at $(q,p)$, we seek open paths  in phase space associated with this point.  The action has three terms, confining the path to low-energy regions of  phase space (more strictly, where the $P$ function of the Hamiltonian is small), minimizing the length of the path and placing the mid-point of the ends of the path  close to the point $(q,p)$.  For points outside the energy surface the dominant path will enclose no area, and the Wigner function will be positive; for points inside the energy surface the path encloses a finite area, and the sign of Wigner function will depend on the area enclosed. 

\appendix \section*{Appendix: Saddle point evaluation of the path integral}
\setcounter{section}{1}
Here we outline the evaluation of the path integral for the Wigner function (\ref{eq:Wsp}) in the saddle-point approximation.  The action $S_L[\gamma,\alpha]$ (\ref{eq:SN}) is an analytic function of $\{\theta_l\}$ (but not of $\{\gamma_l\}$),
\begin{equation}
\label{eq:Squad}
S_L[\gamma,\alpha] = S_L^{(0)}(\alpha) +\frac 1 2 \sum_{lm} S_L^{(2)}(\alpha)_{lm}
(\theta_l-\theta_l^{(0)})(\theta_m-\theta_m^{(0)}) + \ldots,
\end{equation}
where the action is stationary, $\partial S_L/\partial \theta_l = 0,$ at the saddle point $\theta_l  = \theta_l^{(0)}$.  The derivatives are
\begin{eqnarray}
\frac {\partial S_L}{\partial \theta_1}& = & 
\rmi\left[ -r^2\left(\rme^{\rmi(\theta_1-\theta_2)}+\rme^{\rmi(\theta_L-\theta_1)}\right) +2rs \rme^{-\rmi\theta_1}\right], \label{eq:S11}\\
\frac {\partial S_L}{\partial \theta_l}& = & 
\rmi r^2\left(\rme^{\rmi(\theta_{l-1}-\theta_l)}-\rme^{\rmi(\theta_l-\theta_{l+1})}\right), 1<l<L, \label{eq:S1l} \\
\frac {\partial S_L}{\partial \theta_L}& = & 
\rmi\left[ r^2\left(\rme^{\rmi(\theta_{L-1}-\theta_L)}+\rme^{\rmi(\theta_L-\theta_1)}\right) -2rs \rme^{\rmi\theta_L}\right]. \label{eq:S1L}
\end{eqnarray}
Setting the derivative (\ref{eq:S1l}) to zero shows that the angles are equally spaced around an arc of angle $2\theta$,
\begin{equation}
\label{eq:theta0}
\theta_l^{(0)} =\left( \frac{2l-L-1}{L-1} \right) \theta,
\end{equation}
where substitution in equation (\ref{eq:S11}) or (\ref{eq:S1L}) gives an implicit equation for $\theta$:
\begin{equation}
\label{eq:theta}
s = \frac r 2 \left( \rme^{\rmi\theta} + \rme^{-\rmi\theta(L+1)/(L-1)} \right).
\end{equation}
The saddle point has a non-zero imaginary part for finite $L$, and for $s>r$ the saddle-point value of $\theta$ is purely imaginary. However, as $L\rightarrow\infty$ the saddle point becomes
\begin{equation}
\label{eq:thetainf}
s = r\cos\theta,
\end{equation}
so that the chord joining the end-points of the arc passes through the point $\alpha$. 

The action at the saddle point follows from substitution of Eqs.~(\ref{eq:theta0}--\ref{eq:theta}) into equation (\ref{eq:SN}):
\begin{eqnarray}
S_L^{(0)}(\alpha) & = & Lr^2 \left(1-\rme^{-2\rmi\theta/(L-1)}\right) + \frac {r^2} 2  \left( \rme^{-2\rmi\theta(L+1)/(L-1)} -\rme^{2\rmi\theta} \right)\label{eq:SL0}\\
 & = & \rmi r^2\left(2\theta-\sin 2\theta\right) + \Or(L^{-1}) \label{eq:Sinf0} \\
 & = & 2\rmi r^2\left(\cos^{-1}\frac s r-s\sqrt{r^2-s^2}\right) + \Or(L^{-1}). \label{eq:Sinf1}
\end{eqnarray}

The second derivative at the saddle point is given by the $L\times L$ matrix (for $L>2$)
\begin{equation}
\label{eq:S2mat}
S_L^{(2)}(\alpha) = r^2 \rme^{-2\rmi\theta/(L-1)}
\left(
\begin{array} {cccccc} 
2 & -1 & 0 & \cdots & 0 & t \\
-1 & 2 & -1 & \cdots & 0 & 0 \\
0 & -1 & 2 & \cdots & 0 & 0 \\
\vdots & \vdots & \vdots & \ddots & \vdots & \vdots \\
0 & 0 &0 & \cdots & 2 & -1 \\
t& 0 & 0 & \cdots & -1 & 2 
\end{array}
\right),
\end{equation}
where
\begin{equation}
\label{eq:t}
t = \rme^{2\rmi L\theta/(L-1)}.
\end{equation}
The determinant of the matrix is straightforward to evaluate:
\begin{equation}
\label{eq:detS}
\det S_L^{(2)}(\alpha) = r^{2L}t^{-1}\left[(1+L) + 2t + (1-L)t^2\right].
\end{equation}

The Wigner function (\ref{eq:WLexact}) in the saddle-point approximation becomes \cite{Kleinert}
\begin{eqnarray}
\fl W_L^{\mathrm{sp}}(\alpha,\alpha^*;r^2) &=&
\frac {\rme^{-S_L^{(0)}(\alpha)}} {(2\pi)^{L/2} Z_L(r^2)} (\det S_L^{(2)}(\alpha))^{-1/2} \\
& = & \frac {\rme^{\rmi r^2(\sin 2\theta - 2\theta+\Or(L^{-1}))}} {L^{1/2}(2\pi)^{L/2} Z_L(r^2)} r^{-L}(1-e^{4i\theta}+\Or(L^{-1}))^{-1/2}\rme^{\rmi\theta}.
\end{eqnarray}
Substituting for $\theta$ from equation (\ref{eq:thetainf}) gives
\begin{equation}
\label{eq:Wsp1}
W_L^{\mathrm{sp}}(\alpha,\alpha^*;r^2) \approx \frac {\rme^{2\rmi s\sqrt{r^2-s^2}-2\rmi r^2\cos^{-1}(s/r) +\rmi\pi/4}}
{r^{L-1} {2L^{1/2}(2\pi)^{L/2} Z_L(r^2)} \left(s\sqrt{r^2-s^2}\right)^{1/2}}.
\end{equation}
Adding the time-reversed saddle-point $\theta \rightarrow -\theta$ and setting $r^2=n+1/2$ gives our result (\ref{eq:Wsp}) for the number-state Wigner function
\begin{equation}
\label{eq:Wsp2}
\fl W_L^{\mathrm{sp}}(\alpha,\alpha^*;n+\frac12) = \frac 
{\cos\left((2n+1)\cos^{-1} \left(\frac{|\alpha|}{\sqrt{n+\frac12}}\right)-2|\alpha|\sqrt{n+\frac12 - |\alpha|^2}- \pi/4+\Or(L^{-1})\right)}
{(2\pi)^{L/2}L^{1/2}Z_L(n+1/2)\left(|\alpha|^2(n+\frac 12-|\alpha|^2)\right)^{1/4}}.
\end{equation}
As expected, the area enclosed by the path largely determines the phase, with a small phase shift from the saddle-point expansion.

For $s>n+\frac 1 2$ (outside the energy surface) there is one imaginary saddle point,
\begin{equation}
\label{eq:thetaim}
\theta = \rmi \cos^{-1} \frac s r
\end{equation}
and the Wigner function becomes
 \begin{equation}
\label{eq:Wforb}
\fl W_L^{\mathrm{sp}}(\alpha,\alpha^*;n+\frac12) = \frac 
{\exp\left((2n+1)\cosh^{-1} \left(\frac{|\alpha|}{\sqrt{n+\frac12}}\right)-2|\alpha|\sqrt{ |\alpha|^2-n-\frac12}+\Or(L^{-1})\right)}
{2(2\pi)^{L/2}L^{1/2}Z_L(n+\frac12)\left(|\alpha|^2(|\alpha|^2-n-\frac12)\right)^{1/4}}.
\end{equation}

For large $n$ we can use Stirling's approximation for $n! \approx (2\pi)^{1/2}e^{-n+1/12n}n^{n+1/2}$ to obtain
\begin{equation}
\label{eq:ZLStirling}
Z_L(n+1/2) = \left[2\pi\left(n+\frac 5 {12} + \Or(n^{-1})\right)\right]^{-L/2},
\end{equation}
consistent with the standard deviation $\sqrt {n+\frac 1 2}$ of the Poisson distribution (\ref{eq:Poisson}), and hence
\begin{equation}
\label{eq:Wspall}
\fl W_L^{\mathrm{sp}}(\alpha,\alpha^*;n+\frac12) \approx 
\left\{ \begin{array}{ll}
  \frac 
{\cos\left((2n+1)\cos^{-1} \left(\frac{|\alpha|}{\sqrt{n+\frac12}}\right)-2|\alpha|\sqrt{n+1/2 - |\alpha|^2}- \pi/4\right)}
{L^{1/2}\left(1+\frac {1}{24n}\right)^L \left[\frac{|\alpha|^2}{n+\frac 1 2}\left(1-\frac{|\alpha|^2}{n+\frac 1 2}\right)\right]^{1/4}} ,   &  |\alpha|<\sqrt{n+\frac 1 2}  \\ & \\
\frac{\exp\left((2n+1)\cosh^{-1} \left(\frac{|\alpha|}{\sqrt{n+1/2}}\right)-2|\alpha|\sqrt{ |\alpha|^2-n-\frac12}\right)} 
{2L^{1/2}\left(1+\frac {1}{24n}\right)^L \left[\frac{|\alpha|^2}{n+\frac 1 2}\left(\frac{|\alpha|^2}{n+\frac 1 2}-1\right)\right]^{1/4}},   &  |\alpha|>\sqrt{n+\frac 1 2}
\end{array}\right.
.
\end{equation}
The functional form agrees with the WKB expression (\ref{eq:WWKB}) of Berry up to a constant factor weakly dependent on $n$ and $L$ (but divergent in the $L\rightarrow\infty$ limit).
\section*{References}
\bibliography{Wigner2}
\end{document}